# Schizophrenic Representative Investors

Philip Z. Maymin

*Department of Finance and Risk Engineering, NYU-Polytechnic Institute, Brooklyn, NY*

Philip Z. Maymin
Department of Finance and Risk Engineering
NYU-Polytechnic Institute
Six MetroTech Center
Brooklyn, NY 11201

Phone: (718)260-3175
Fax: (718)260-3355

Email: phil@maymin.com

Representative investors whose behaviour is modelled by a deterministic finite automaton generate complexity both in the time series of each asset and in the cross-sectional correlation when the rule governing their behaviour is schizophrenic, meaning the investor must hold multiple seemingly contradictory beliefs simultaneously, either by switching between two different rules at each time step, or computing different responses to different assets.

Keywords: complexity; representative investor; deterministic; behavioural

# Schizophrenic Representative Investors

**Abstract:** Representative investors whose behavior is modeled by a deterministic finite automaton generate complexity both in the time series of each asset and in the cross-sectional correlation when the rule governing their behavior is schizophrenic, meaning the investor holds multiple seemingly contradictory beliefs simultaneously, either by switching between two different rules at each time step, or computing different responses to different assets.

## 1. Introduction

Observed complexity need not be the result of a complex underlying process. Most famously, Conway's Game of Life has very simple rules for the evolution of cells on an infinite checkerboard, but generates very complex behavior (Gardner 1970). Indeed, simple rules often lead not merely to complex-looking behavior, but to complete universal computability (Wolfram 2005); in this sense, simple rules can lead to maximally complex behavior. It is therefore useful to explore the simplest possible rules that can lead to complexity.

A recent paper introduced a model of a deterministic representative agent trading a single asset based solely on its price history and found that of the 128 distinct possible rules, only one generated complexity. Because the rule is unique, and because it relies on only one investor trading only one asset based only on past movements, this model is known as the minimal model of financial complexity (Maymin 2011).

This paper extends the minimal model of financial complexity in two ways. The first extension ostensibly allows for multiple investors by allowing different rules to govern on consecutive time steps. One interpretation of this wrinkle is that there are two investors who take turns being the "representative" investor. However, an alternative interpretation is that there is still a single unique representative investor, but that investor follows a schizophrenic rule that is a combination of two other rules. The impact of this first extension is that there are many pairs of rules, each of which by themselves do not generate complexity, which do generate complexity when they are alternated.

The second extension allows for multiple assets. The representative agent follows a single deterministic trading rule, but that rule governs the portfolio decision of every asset jointly. In other words, rather than merely deciding at each time step whether to buy or sell the market asset based on the past few movements of the market asset, the representative investor must decide at each time step whether to buy or sell each of the available assets based on the past few movements of each of the assets. This representative investor is considered schizophrenic as well because he essentially evaluates assets differently even if they have the same history of past movements.

The rest of this paper is organized as follows. Section 2 introduces the terminology and reviews the minimal model of financial complexity. Section 3 demonstrates the results of the schizophrenic investor who switches rules each time step. Section 4 demonstrates the results of the schizophrenic investor who follows a single rule but on two assets. Section 5 concludes.

## 2. The Basic Model

The basic model uses an iterated finite automaton (IFA) applied to the past *w* movements of the asset in reverse order (i..e, starting with the most recent one first). An IFA is completely represented by a state transition diagram. For example, the minimal model of financial complexity known as rule 54 follows the following diagram:

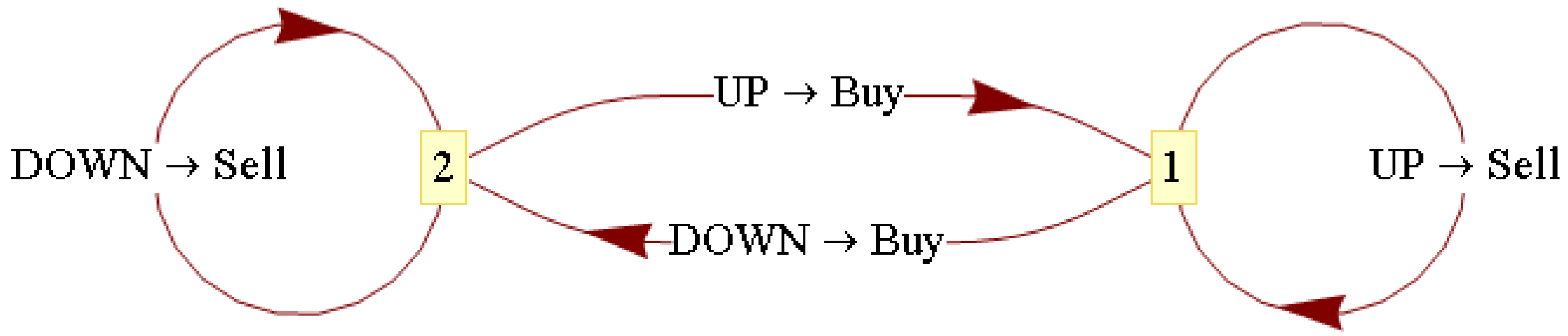

Every IFA has a finite number of states. The minimal number of states for non-degenerate cases turns out to be two; these two states are labeled "1" and "2" in the above diagram. The investor always started in the special state "1" each day. Then, he looks back on yesterday's movement. Was the market UP yesterday? Then he follows the arrow leaving state "1" that is labeled "UP." That arrow leads back to state "1." Was the market UP the day before yesterday? Then the investor again follows the "UP" arrow. If the market had been down, he would have followed the "DOWN" arrow.

The right hand side of the label of each arrow that he is following represents the investor's current inclination to buy or sell the asset. So after looking at yesterday's UP return, he would be inclined to sell, but he does not do so yet, because he has not gone through all of the past price movements. Only on the last movement does the investor then decide whether to buy or sell, based on what is on the most recent arrow that he followed.

The number of past movements that the investor looks at is called his lookback window and is denoted $w$. The number of states in his IFA is denoted $s$ and the number of possible "actions" is denoted by $k$. An action is a possible decision he might make. In the example above, the investor had $k = 2$ possible actions, buy or sell. If the investor could also choose to hold, then he would have $k = 3$ possible actions.

The initial history is for convenience assumed to be a sequence of $w$ buys. For rules that generate complexity, this choice of initial history is arbitrary because any other choice merely shifts the entire time series forward or backward.

The number of different possible rules for an IFA with $s$ states and $k$ actions is $(s \cdot k)^{s \cdot k}$ and (Wolfram 2003) lays out a convenient numbering scheme such that any such IFA can be uniquely identified by a number between 0 and $(s \cdot k)^{s \cdot k} - 1$. For example, in the 2-state, 2-action minimal IFAs, there are $4^4 = 256$ different possible rules. However, half of those rules are effectively duplicates because they merely relabel state "1" as state "2" and vice-versa. Of the 128 unique rules, only one (rule 54) generates complex behavior, where complexity is defined as having a period of at least half of the maximum possible cycle length: for a lookback window of $w$ and $k$ possible actions, the IFA must cycle within $k^w$ time steps because the sequence of past history must have repeated itself at least once by then.

## 3. Alternating Rules

Imagine if the representative investor changes which rule he follows every day. Specifically imagine if he alternates between two different rules. On day one and all subsequent odd-numbered days, he evaluates the last $w$ price movements of the market asset by following rule 1. On day two and all subsequent even-numbered days, he evaluates the last $w$ price movements of the market asset by following rule 2. In this sense he is a schizophrenic representative investor, where schizophrenic is used in its non-medical sense to mean a person simultaneously holding conflicting beliefs.

Would such a situation result in more rules that generate complexity? This question demonstrates the necessity of actually doing the simulation to determine the answer. For example, a slightly different question generates a completely different answer: if we fix the rule but alternate the lookback window between $w$ to $w+1$, it turns out there is no rule, including rule 54, which generates complexity.

However, in the case of alternating rules, there are substantially more combinations that generate complexity. The table below lists all of the pairs of rules that generate complex financial time series with a period at least as long as that of the single rule 54.

| | **39** | **52** | **54** | **60** | **97** | **99** | **114** | **141** | **148** | **156** | **188** | **193** | **195** | **201** | **216** | **227** | **233** |
|---|---|---|---|---|---|---|---|---|---|---|---|---|---|---|---|---|---|
| **39** | | | 1022 | | | 1022 | | | | 1022 | | | | 1022 | | | |
| **52** | | | | | 1274 | 1588 | | | | 1588 | 1272 | | | | | | |
| **54** | | | 889 | | 1392 | 1778 | 1022 | 1022 | 1588 | 1778 | 1392 | | | 889 | | | |
| **60** | | | | | | | | | | | | | | | | | |
| **97** | | 1274 | 1392 | | | | | | | | | | | 1392 | | | 1284 |
| **99** | 1022 | 1588 | 1778 | | | | | | | | | | | 1778 | 1022 | | |
| **114** | | | 1022 | | | | | | | | | | | 1022 | | | |
| **141** | | | 1022 | | | | | | | | | | | 1022 | | | |
| **148** | | | 1588 | | | | | | | | | 1274 | | 1588 | | | |
| **156** | 1022 | 1588 | 1778 | | | | | | | | | | | 1778 | 1022 | | |
| **188** | | 1272 | 1392 | | | | | | | | | | | 1392 | | | 1272 |
| **193** | | | | | | | | | 1274 | | | | | | | 1284 | |
| **195** | | | | | | | | | | | | | | | | | |
| **201** | | | 889 | | 1392 | 1778 | 1022 | 1022 | 1588 | 1778 | 1392 | | | 889 | | | |
| **216** | | | 1022 | | | 1022 | | | | 1022 | | | | 1022 | | | |
| **227** | | | 1496 | | | | | | | | | 1284 | | 1496 | | | |
| **233** | | | | | 1284 | | | | | | 1272 | | | | | | |

There are a total of 68 rule pairs exhibiting complexity and only two singlet rules: the rule pair (54, 54), which is the same as the single rule 54, and the rule pair (201, 201), which is the same as the single rule 201. Each has a period of 889. Recall that rule 201 is the same as rule 54 but with states 1 and 2 relabeled.

The rule pairs are not symmetric. Notice that the rule pair (39, 54) is complex but the rule pair (54, 39) is not.

The figure below shows an example of how two rules, each of which by itself generates repetitive, non-complex time series, can combine to generate extended complexity.

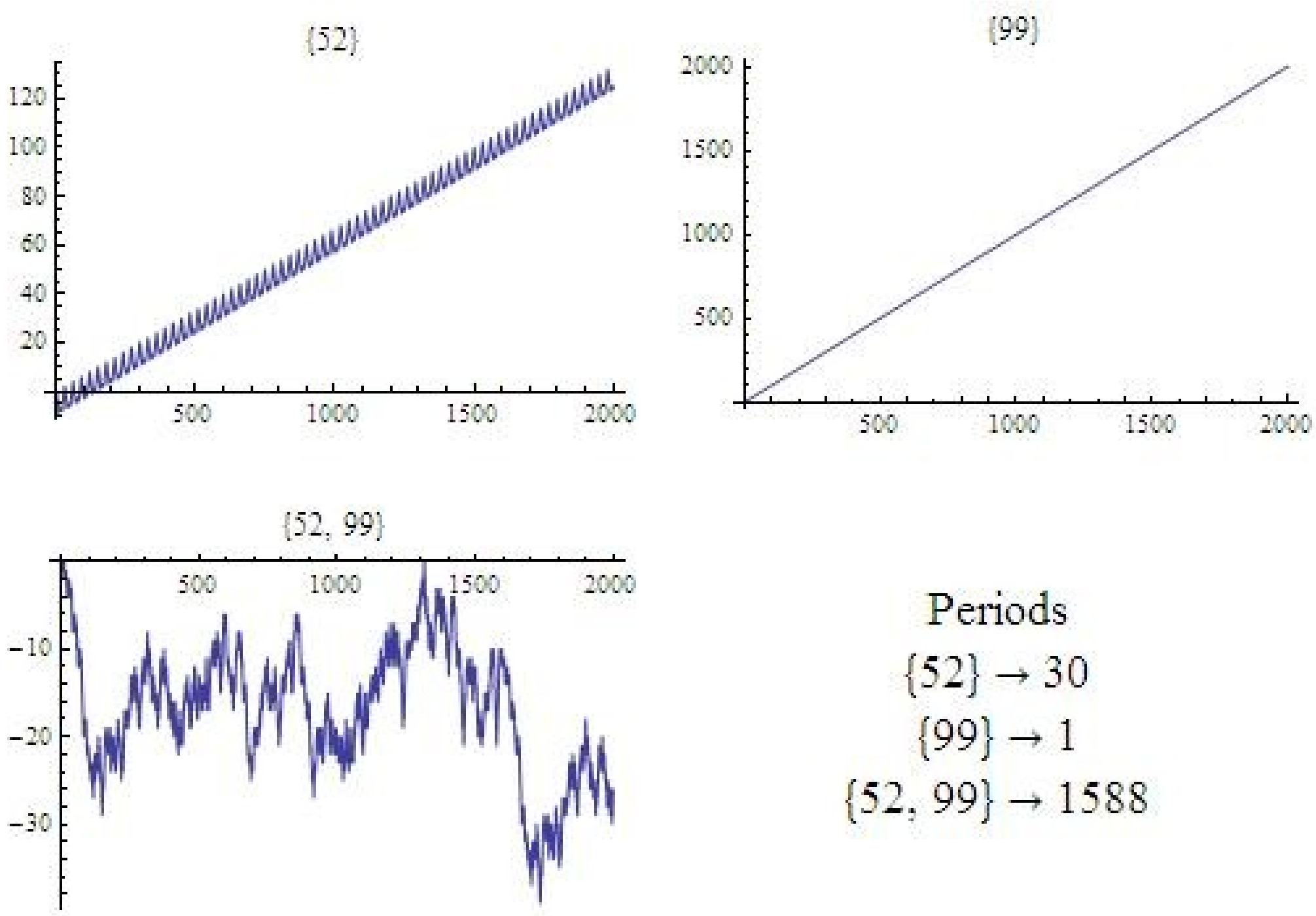

Rule 52 by itself has a period of 30, meaning its sequence of UPs and DOWNs repeat every 30 time steps. Rule 99 is even simpler: it always goes UP so it has a period of 1 time step. But when the schizophrenic representative investor alternates between rule 52 and rule 99, the resulting time series is even more complex than the single rule 54: its period is 1,588.

The figure below graphs all of the distinct complex time series that can result from the rule pairs. A schizophrenic representative investor has substantially more variety and diversity in complexity than the single rule 54.

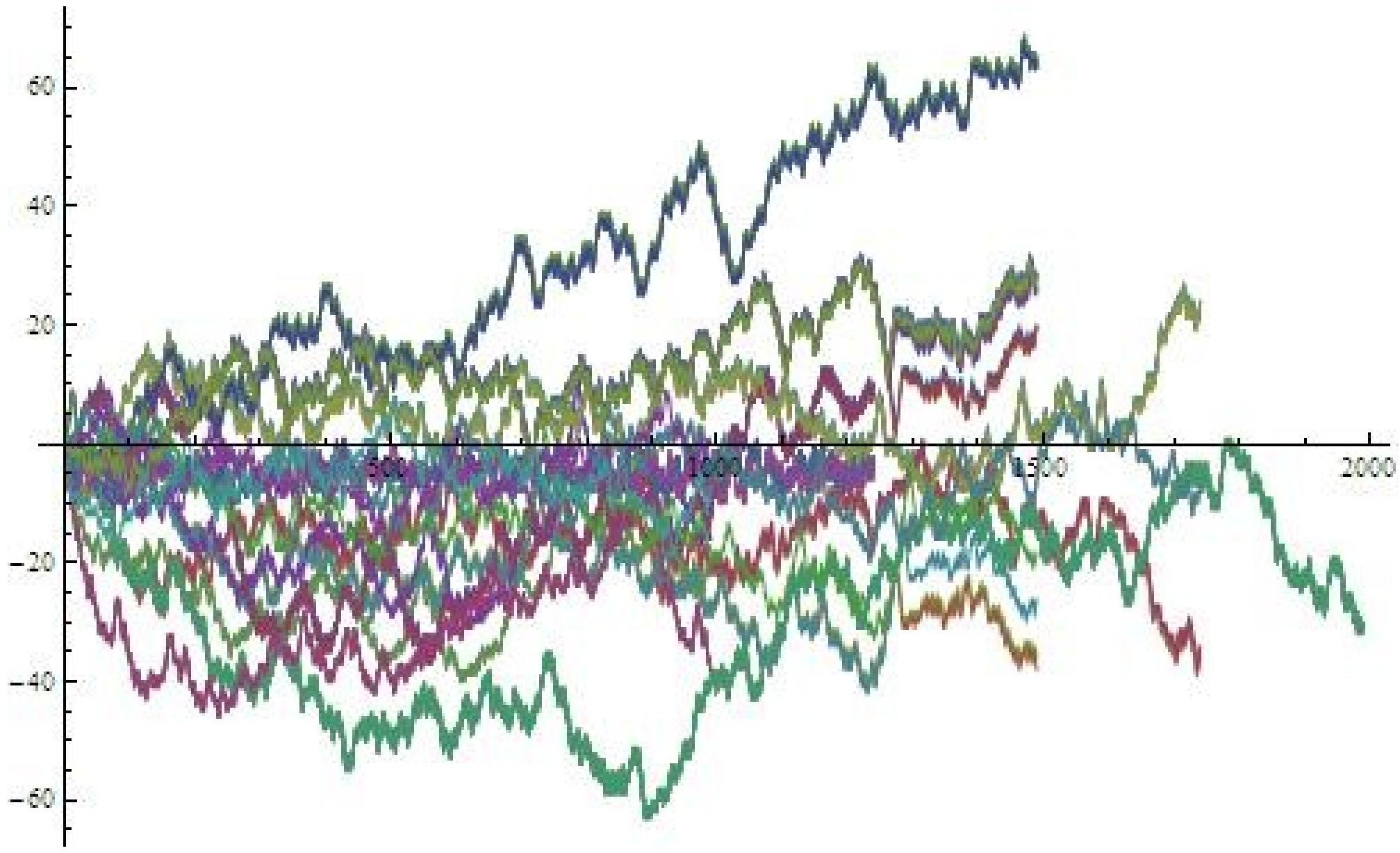

## 4. Multiple Assets

To extend the basic model to allow for the simultaneous pricing of multiple assets, we can interpret each of the $k$ actions as representing an entire portfolio. Consider the case of $m = 2$ assets, each of which could be either bought or sold. Then there are $k = 2^m$ possible actions. Interpreting the number $k$ in base-2 gives us a zero or a one for each asset, representing either a sale or a buy, respectively.

Allowing multiple assets increases the search space of possible rules exponentially because the number of possible rules is $(s \cdot 2^m)^{s \cdot 2^m}$. Even for $m = 2$ assets and still only $s = 2$ states, there are $8^8 = 16{,}777{,}216$ possible rules.

The natural extension of assuming all up movements for an initial history is to assume that each asset had all up movements in its initial history; thus, the initial history would be a $w$-length sequence of identical actions, namely $2^m - 1$, which in binomial notion is a sequence of $m$ ones.

How many of those millions of rules actually generate complex behavior in each of the financial assets? We can run each of the rules with a lookback window of e.g. $w = 12$ and count the number of rules that generate asset histories for 100 time steps without repetition in either asset and for which the two assets are not either identical (correlation of one) or exactly opposite (correlation of negative one).

There are 6,266 rules that fit those criteria. Some of those rules are repeats; there are only 3,986 distinct rules. That means less than 2 percent of 1 percent of all of the 16,777,216 possible rules generate complex behavior. Unlike the minimal model of financial complexity with a single asset, these rules are not unique, but they are still quite rare.

What do the evolutions of these rules look like? The figure below shows the accumulation of 500 time steps for both assets from each of the 3,986 complex rules. Thus, there are 7,972 time series plots below. Note that the most extreme possible time series would be lines starting from the origin and ending either at +500 or -500 because those would represent all UPs or all DOWNs for the asset in question. None such appear because such evolutions are non-complex and were filtered when we discarded short-cycle evolutions.

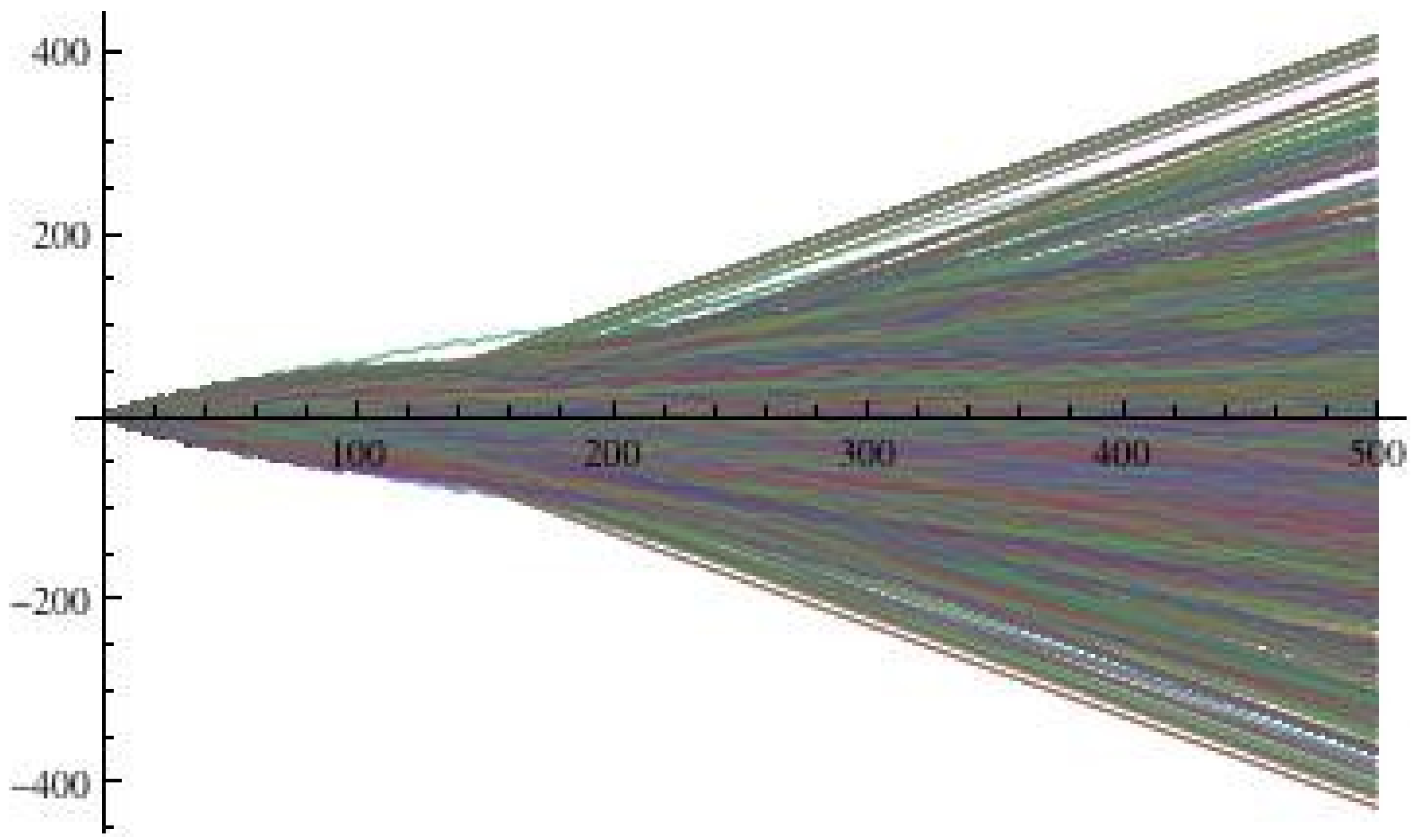

Just about every possible path seems to be represented here. But with what probability? The graph below shows the histogram of the terminal values of each of the assets.

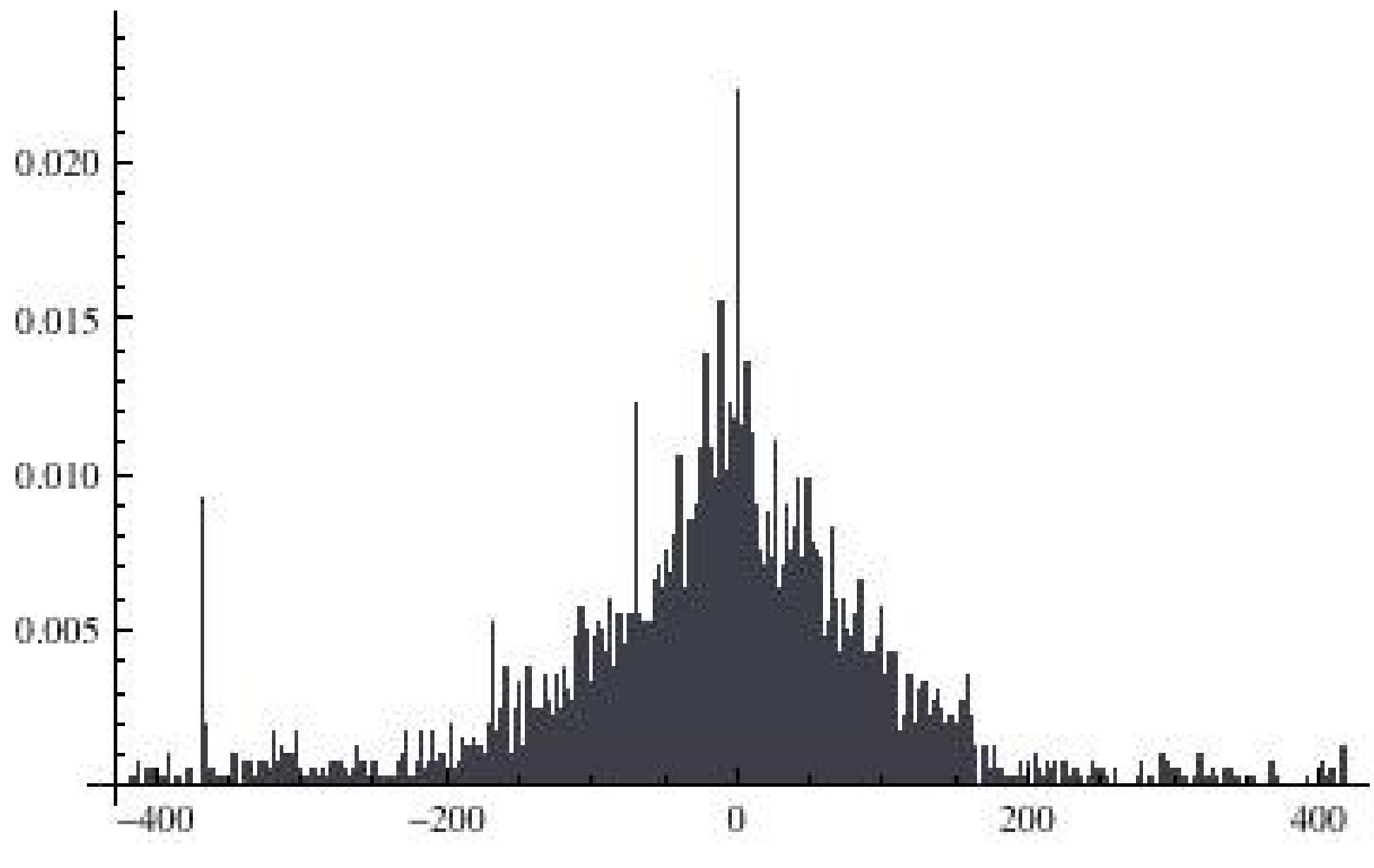

Notice that the distribution appears to be slightly negative skewed and fat-tailed. Indeed, the skewness of the distribution is -0.39 and the kurtosis is 5.2. This conforms with typical stylized facts about the markets which also often appear to have negative skewness and excess kurtosis.

However, those numbers are cross-sectional in nature, and look across a variety of different rules. A more interesting measure of complexity is not the simple overall value but the diversity that is possible.

What does the distribution of the skewness and kurtosis look like if we compute them for each possible path of each asset? The figures below show the results. Note that these histograms are plotted on a log-probability scale so that the outliers are visible.

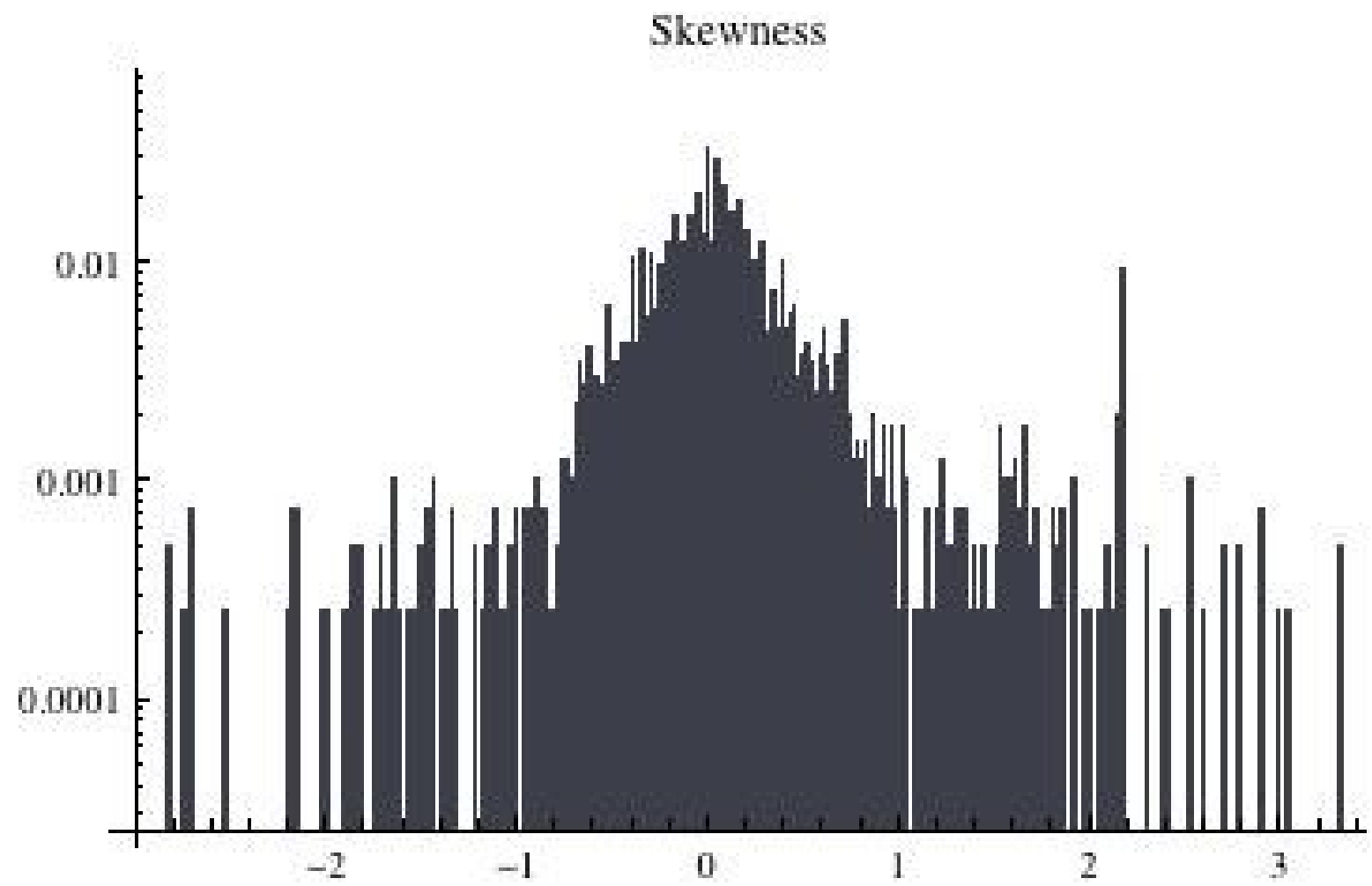

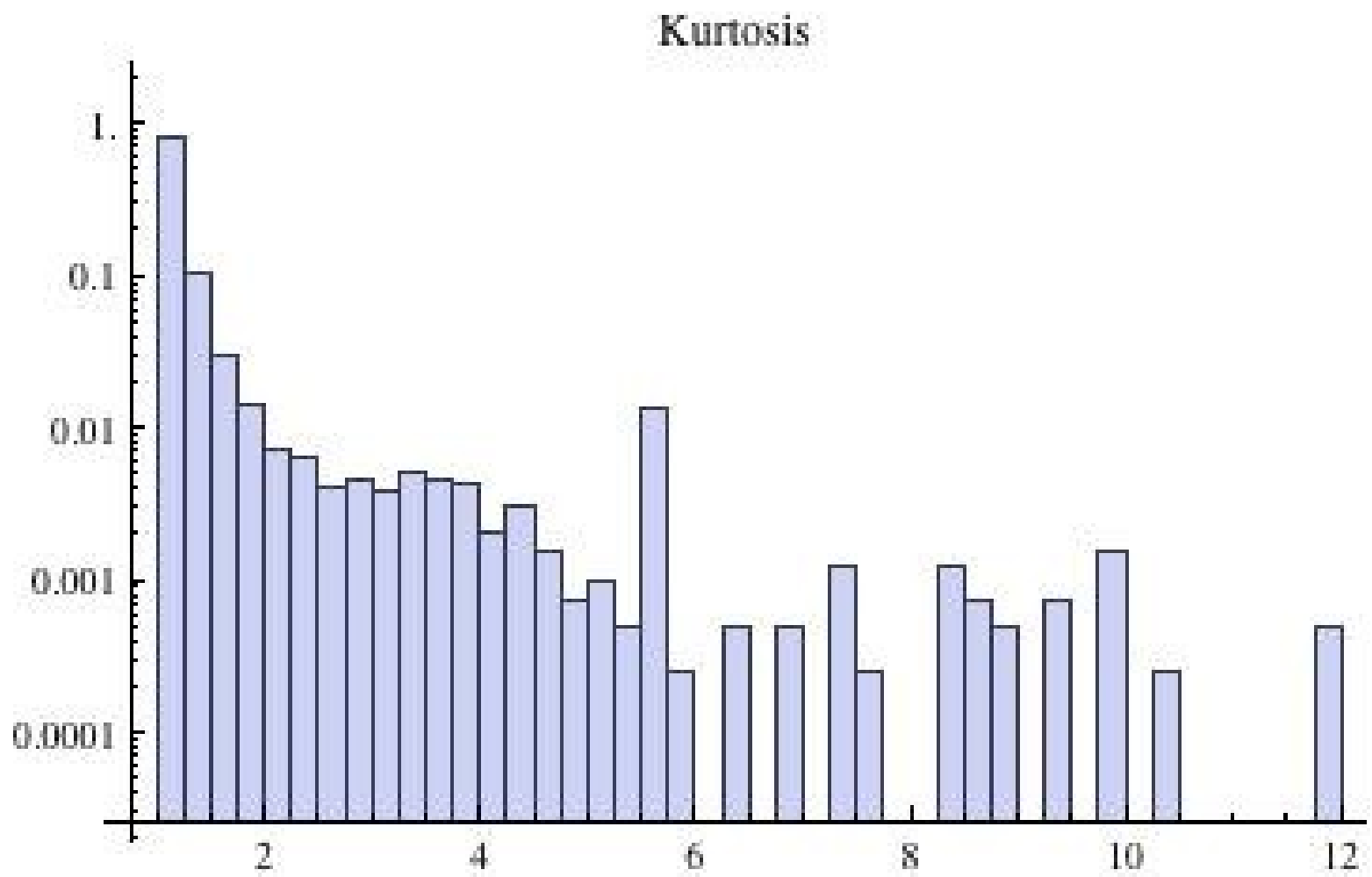

Notice the broad range of skewness and kurtosis that is possible. Just as in the real world we see both positive and negative skewness for different assets, so too do we see them as the result of this model. And just as in the real world we see occasionally thinner tails (kurtosis below three), medium tails (kurtosis around three), and fat tails (kurtosis above three), so too do we see them emerge from this model, including outliers as high as 12.

We can also look at the distribution of pairwise correlations by calculating the correlation between the two assets in a given rule, and plotting the histogram of all such correlations across all of the interesting rules.

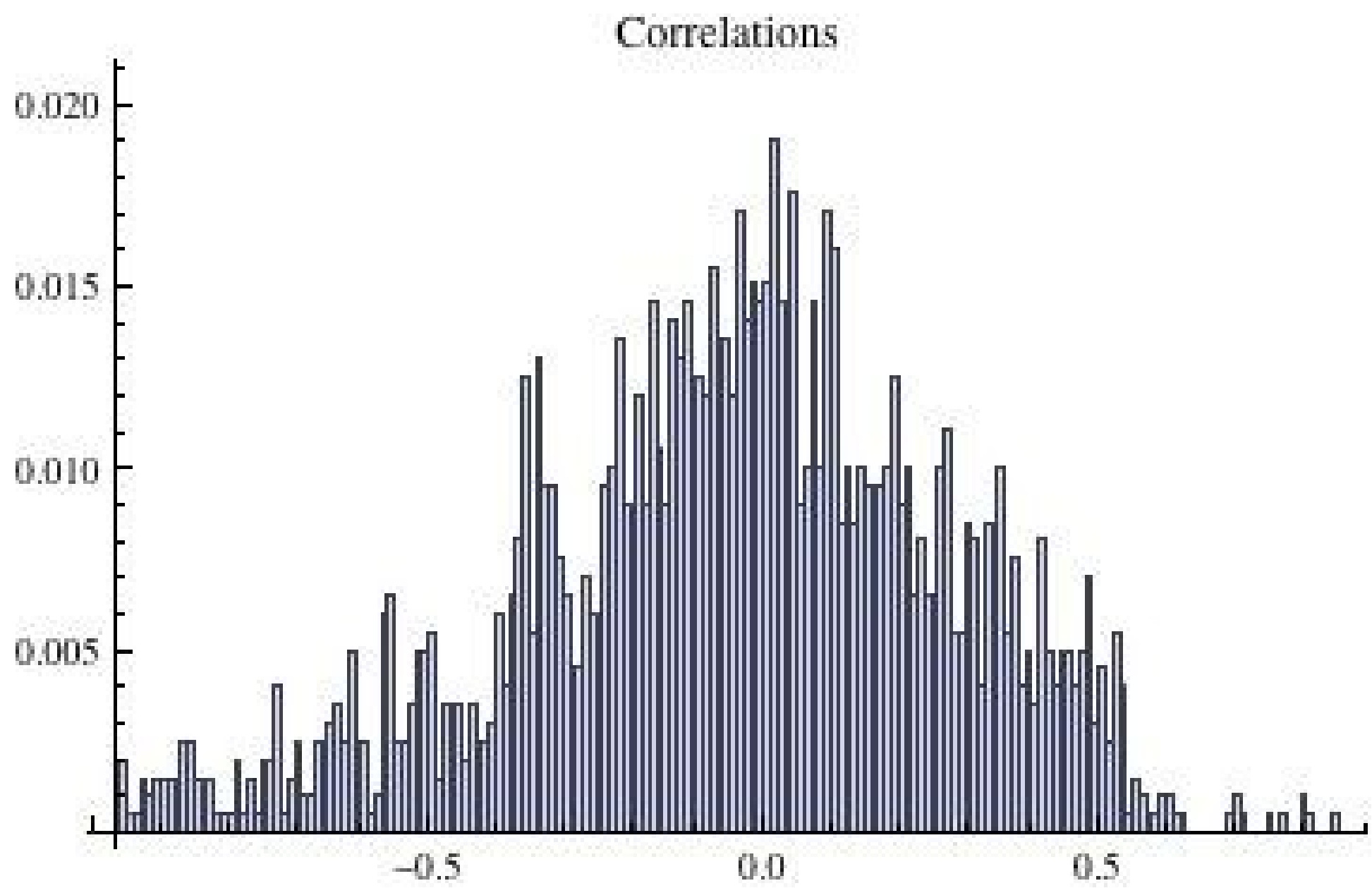

The pairwise correlations tend to be clustered around zero though there are extremes stretching as far as -0.96 and +0.86. But how do these correlations relate to recent activity in the two assets?

Let's calculate the rolling correlation over 50 time steps and compare it to the return of the best performing asset in that same period of 50 time steps. Why the return of the better performing asset? Because we know if the better one has a negative return, then so must the other one, and because this way we can see if there is a difference between environments where one asset is up and one is down as opposed to an environment where both are down. With those numbers, we can then plot the relationship between the correlation and the return of the better asset across all time periods and all interesting rules. That figure is below.

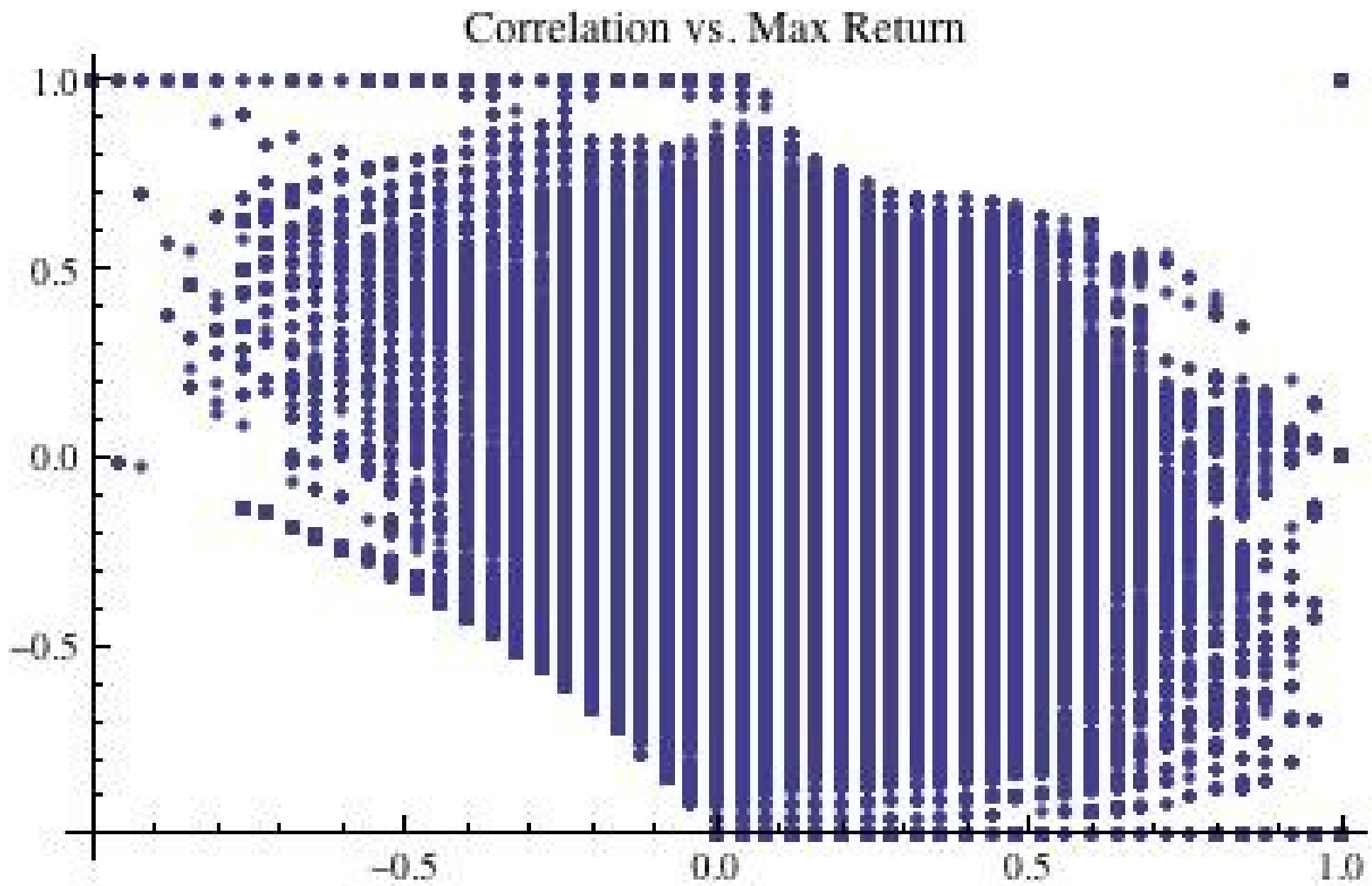

Notice that when the better performing asset is negative, meaning both assets had negative returns for the period, the correlation tends to increase the worse the performance is. In other words, in times of broad market corrections, the deeper the correction, the higher the correlation. This also conforms with a broad stylized fact about the market.

Similarly, but less pronounced, as the better performing asset increases in return, the correlation has a tendency to decrease.

Combining these two insights and recalling that the implied volatility for options on an index depends strongly on the implied correlation between the constituent assets comprising the index suggests that the implied volatility skews on options traded on the market would have a steep skew for puts and a less steep but still downward sloping skew on calls, matching another stylized fact about index implied volatility markets.

## 5. Conclusion and Summary

The complexity we observe in financial time series need not result from complexity in the fundamental rules of the market. A generic description of simple rules is presented that generates complexity similar to those in the real world. Even a representative investor trading a single market asset can generate complexity with a simple, naïve, and deterministic rule.

Multiple traders can be modeled as a sequence of representative investors each following a simple rule. An alternative interpretation is that there is still only a single representative investor, but the rule he follows alternates between days.

Multiple assets can be modeled by reinterpreting the possible actions of each trade to represent choices about every possible asset.

The variety and complexity of possible results matches that found in the real world. The generated time series tend to have negative skewness, high kurtosis, and correlations that increase during market downturns.

All these results occur without any parameter fitting but merely by exploring the space of possible rules given a simple framework.